# Patterning of diamond like carbon films for sensor applications using silicon containing thermoplastic resist (SiPol) as a hard mask

D. Virganavičius [a,b], V.J. Cadarso [a], R. Kirchner [a], L. Stankevičius [b], T. Tamulevičius [b], S. Tamulevičius [b], H. Schift [a]

[a] Paul Scherrer Institut, Laboratory of Micro- and Nanotechnology, 5232 Villigen PSI, Switzerland

[b] Kaunas University of Technology, Institute of Materials Science, 51423 Kaunas, Lithuania

e-mail: helmut.schift@psi.ch

Keywords: diamond like carbon, nanocomposites, thermal nanoimprint lithography, plasma etching, refractive index sensing

## Abstract

Patterning of diamond-like carbon (DLC) and DLC:metal nanocomposites is of interest for an increasing number of applications. We demonstrate a nanoimprint lithography process based on silicon containing thermoplastic resist combined with plasma etching for straightforward patterning of such films. A variety of different structures with few hundred nanometer feature size and moderate aspect ratios were successfully realized. The quality of produced patterns was directly investigated by the means of optical and scanning electron microscopy (SEM). Such structures were further assessed by employing them in the development of gratings for guided mode resonance (GMR) effect. Optical characterization of such leaky waveguide was compared with numerical simulations based on rigorous coupled wave analysis method with good agreement. The use of such structures as refractive index variation sensors is demonstrated with sensitivity up to 319 nm/RIU, achieving an improvement close to 450% in sensitivity compared to previously reported similar sensors. This pronounced GMR signal fully validates the employed DLC material, the technology to pattern it and the possibility to develop DLC based gratings as corrosion and wear resistant refractometry sensors that are able to operate under harsh conditions providing great value and versatility.

1. Introduction

Patterning of diamond and diamond-like materials is of interest for a number of applications, such as stamps in nanoimprint lithography (NIL), in hard X-ray optics or infrared optics [1-3], due to their unique properties (i.e. high hardness, optical transparence, high refractive index, and chemical inertness). Particularly diamond-like carbon (DLC) films have become attractive because of their physical and mechanical properties similar to those of diamond, cost-efficient fabrication and room temperature deposition. DLC is an amorphous material consisting of $sp^3$ and $sp^2$ type bonded carbon atoms with the possible presence of hydrogen and other dopants including silicon dioxide, metals, etc. There exists a variety of amorphous carbon materials that can be fabricated employing different deposition techniques using a vast number of precursors. Therefore optical, electrical, mechanical and other properties of DLC films can also be tuned greatly. More detailed information on diamond like coatings can be found in review articles [4, 5].

Despite the great potential of such materials, their patterning is still challenging and usually depends on complex processing involving multilayer masks and elaborate etching schemes [6-9], relies on costly low throughput patterning [10], or has very limited feasibility [11]. However, well-defined DLC micro- and nanostructures are highly desirable for variety of optical and MEMS devices [12-14]. In particular, DLC based guided mode resonance (GMR) structures appear to be very attractive sensing platforms. The GMR phenomenon occurs in subwavelength dielectric periodic structures, where under resonant conditions incident light can be coupled to leaky surface wave and guided by the nanostructure resulting in Wood-type anomalies at the resonant wavelength in the transmission or reflection spectra. Resonant spectral response and wavelength depends on material and geometry of periodic structure. Moreover resonance conditions are also very sensitive to the effective refractive index and extinction changes in the vicinity of the grating surface, i.e. evanescent field of the leaky wave. Combination of GMR sensor together with fluid cell enables to use the leaky waveguide chip as an in-situ refractive index sensor with sensitivities down to $10^{-6}$ RIU (refractive index units). Compared to the metallic nanostructures based on localized surface plasmon resonance (LSPR) phenomenon, GMR sensors can offer comparable sensitivity, have the advantage of tunability, which allows working in desired region of the spectrum, and also usually have narrower peaks resulting in orders of magnitude higher figure of merit (FOM) value. Furthermore, it has been demonstrated that GMR sensors can be integrated in lab-on-chip devices [15]. DLC is a promising material for GMR sensors in the VIS-NIR range enabling development of robust wear and corrosion resistant GMR sensors [16, 17]. The biggest shortcoming of DLC is its high internal stress that affects film adhesion to substrate and limits film thickness.

During the growth of the DLC film it is possible to dope it with nanometer scale clusters of metals and semiconductors. This doping of DLC films is an additional advantage since it further broadens their application spectrum and helps to reduce internal stress. However, this also increases the complexity of patterning of such composite materials. Metal-doped DLC has all the intrinsic properties associated with DLC. In addition, DLC films containing metal nanoparticles show interesting optical characteristics due to localized surface plasmon resonance. Silver, gold and copper doped films are particularly interesting because they have a matrix and metal dielectric function and nanoparticle size dependent pronounced plasmonic absorption peak in the visible region of the spectrum. It has been shown that by changing film deposition conditions and applying plasma post processing it is possible to control size and amount of nanoparticles on the surface [18]. Gratings with silver nanoparticles on the surface have the potential to be utilized in sensor devices enabling simultaneous guided-mode and plasmonic resonance based sensing approaches. In this article we refer to hydrogenated amorphous carbon coatings (a-C:H) as DLC and hydrogenated amorphous carbon coatings doped with silver (a-C:H:Ag) as DLC:Ag.

Although DLC films can be effectively etched with oxygen plasma, the use of standard resist (i.e. PMMA, photoresist, etc.) is hampered since the selectivity between typical organic resists and DLC is extremely low [11]. Additionally, the inclusion of the metallic particles (in case of DLC nanocomposites) prohibits the use of metallic hard masks for enhancing selectivity, since it may result in the removal or modification of the nanoparticles on the DLC surface. This renders in an increased complexity for the patterning of DLC films, especially at sub-micron resolutions.

In this work we present a method capable of pattern DLC films in a straightforward way by using thermal nanoimprint lithography (T-NIL) [19] with the silicon containing SiPol resist (micro resist technology GmbH) [20]. The SiPol standard process was optimized, further simplifying the nanoimprint steps for a pattern transfer in DLC using it as a hard mask. This enabled us to pattern both pure DLC and DLC nanocomposite doped with silver nanoparticles in a high throughput, cost-efficient and reproducible way. The obtained high quality DLC structures were used in the development of a GMR refractive index sensor, demonstrating its feasibility for applications by testing with different concentration of organic solvents with known optical properties.

## 2. Materials and Methods

DLC as well as other carbon based materials e.g., diamond or polymeric resists, can be etched by reactive ion etching (RIE) using pure oxygen plasma [3]. As masking layers, it is possible to use polymer resists of suitable height (for "competitive" etching), or hard masks such as thin Cr or $SiO_2$ films for enhanced etch resistance. Often multilayer-resist strategies are pursued, e.g., for patterning a thin hard mask with a polymer resist or by patterning a thick polymeric transfer layer with a thin hard mask. As suitable masking layers we aimed using a thermal NIL (T-NIL) compatible process which would require a minimum processing steps and at the same time offer good selectivity towards DLC enabling direct pattern transfer. Therefore, we chose SiPol (Micro Resist Technology GmbH), a thermoplastic resist (glass transition temperature $T_g$ 63°C) with 10% content of covalently bonded silicon that makes it highly resistant to oxygen plasma. This enables to use it as spin-coatable and imprintable hard mask combined with good etching selectivity towards DLC. Initially SiPol was developed to be used in bilayer system with an organic transfer layer UL1 (Fig. 1 a-b) [20]. In a standard process, the patterned SiPol layer is thinned down by fluorine based etching chemistry, thus etching both the organic and inorganic content of SiPol, which results in removal of the residual layer. After opening the SiPol mask windows, oxygen plasma is used to pattern the purely polymeric transfer layer UL1 and at the same time converting the SiPol masking layer into almost pure silicon oxide. This transfer layer has at the same time the function of improving adhesion to, both substrate and SiPol layer. However, for the etching of DLC materials in pure $O_2$-plasma, the anisotropy is too high for UL1 even when inductively coupled plasma (ICP) systems are used and it is only possible to obtain very shallow DLC structures (Fig. 1b). In order to prevent this, the T-NIL process was optimized as described below to replicate the SiPol directly on the DLC coatings, i.e. without UL1 as intermediate layer resulting in much deeper structures (Fig. 1c). Furthermore, in order to simplify the process and reduce the number of steps, it was intended to eliminate the fluorine chemistry step and use directly the oxygen plasma for conversion of the SiPol into a $SiO_2$ hard mask. For this purpose, the residual layer had to be as thin as possible and ideally close to 0 nm. The optimized processing scheme used to pattern DLC and DLC:Ag films in the step by step fashion is graphically depicted in Fig. 2.

### 2.1. DLC films deposition

DLC films were synthesized on silicon and Borofloat glass substrates by a direct ion beam deposition employing a closed drift ion beam source and acetylene gas as a precursor. Film growth parameters were as following: precursor gas $C_2H_2$ 99.6% purity; base pressure: $2·10^{-4}$ Pa, working pressure $1-2·10^{-2}$ Pa, ion energy 800 eV±20eV, substrate temperature 293±3 K, film thickness 250-450 nm. The refractive index of the films measured by a laser ellipsometer (Gaertner L115) at 633 nm wavelength was 2.8 ± 0.1.

DLC:Ag films were fabricated by direct current reactive magnetron sputtering using an argon and acetylene gas mixture and a silver cathode. Film growth parameters using $C_2H_2$ with 99.6% purity and Ar with 99.9% purity as precursor gases were as following: base pressure: $2·10^{-4}$ Pa, working pressure $1-2·10^{-2}$ Pa. The Ar flow was 70 sccm, $C_2H_2$ flow was 18 sccm, magnetron current 0.1-0.15A, substrate temperature 293±3 K, film thickness 250-450 nm.

Before the deposition, substrates were cleaned in acetone (VLSI quality) in an ultrasound bath for 1 min and rinsed in isopropanol. Additionally, oxygen RIE plasma cleaning was used to remove remaining solvents and activate the surface prior DLC deposition. The Vickers hardness (HV) of fabricated coatings measured by nanoindentation employing a HM2000S (Fischer Technology) tester was 9750

HV for DLC and 7470 HV for DLC:Ag, respectively. More details on the used deposition processes, optical and mechanical properties of the films can be found in [21,22]

## 2.2 Master fabrication

Master stamps with different geometries for nanoimprint process were fabricated using an EBPG 5000Plus (Vistec) direct writing electron beam lithography (EBL) tool. Zeon ZEP520A spin-coated on silicon substrates was used as a positive e-beam resist. After development and etching with oxygen plasma to remove residual layer, the resist was used as a mask to transfer pattern into silicon by $SF_6/C_4F_8$ ICP plasma. Etched Si masters were cleaned with oxygen plasma. Prior to use the Si masters were cleaned by oxygen plasma (Oxford RIE 80 plasma system, 100 W, 1 mbar, 120 s) and then silanized by trichloro-(1H,1H,2H,2H-perfluorooctyl)-silane. Silanization was performed in a custom vacuum chamber at 10 mbar pressure for 10 min. Typical masters were 20×20 mm$^2$ in area and had 250-350 nm deep cavities.

## 2.3 Patterning by thermal nanoimprint lithography

Thermal NIL (T-NIL) was chosen for patterning the resist, due to its large area, high throughput patterning capabilities with nanoscale resolution [23]. Silicon and Borofloat glass substrates with the DLC films were spin-coated with SiPol thermoplastic resist. Borofloat glass is suitable for optical transmission measurements of the GMR signal in gratings. Also it has a similar thermal expansion as silicon helping to minimize pattern displacement and distortion effects arising due to difference of thermal expansion coefficients between template and substrate. A HEX03 (Jenoptik) hot embossing tool was used for imprints. SiPol resist was diluted with propylene glycol methyl ether acetate (PGMEA) to achieve a desired resist thickness ranging from 60 nm to 100 nm depending on stamp pattern. Due to poor SiPol adhesion to the DLC surface, the standard imprint process had to be modified in order to avoid resist peel-off during demolding. An incomplete filling of the stamp cavities strategy was pursued in order to minimize the contact area of resist with the stamp and to achieve zero residual layer (i.e. a completely dewetted surface below the stamp protrusions). Stamps with 250 nm deep patterns with 50% area fill factor were used, with an intended filling height of 50 nm for a 100 nm thick initial resist (calculation with a grating of 400 nm period and 200 nm structure width). Other structures such as hexagonal pillar and hole structures with 300 nm diameter were used, too. The process with the T-NIL was optimized at low temperature (83°C) to avoid other issues such as lack of adhesion, capillary effects or dewetting, since both DLC and SiPol alone are only exhibiting moderate adhesion. Low imprint temperature is an additional advantage of SiPol because heating of hydrogenated DLC films causes changes in dimensions and properties of the material. Additionally, the lower imprint temperature of SiPol (83 °C) instead of PMMA (180 °C) is beneficial as it significantly reduces the tendency of the material to develop capillary bridges.

Imprints with 2-5 min long thermocycles with an applied pressure of $10^7$ Pa·s were found to be sufficient for the resist to homogeneously fill stamp cavities for proper and effective pattern transfer. After cooling down to a temperature 20 °C below the $T_g$ of the polymeric resist material, the stamp was separated from the substrate with the patterned resist layer.

### 2.4 Etching of DLC films

The imprinted SiPol resist, exhibiting a negligible residual layer, was directly used as a mask for pattern transfer into DLC, avoiding the usual window opening with fluorine based plasma. Hence, direct etching of DLC with pure oxygen plasma can be performed immediately after imprint. When etched by pure oxygen, the organic content is removed at the surface and SiPol converts into silicon oxide which has high etch resistance. In contrast to the standard process presented before, it converts only a thin surface skin into $SiO_2$ and thus the remaining SiPol is still organic in the bulk with negligible further thinning down.

As for etching of DLC process is fairly straightforward. The exposed DLC surface is chemically etched by reactive oxygen species creating volatile products in combination with physical etching caused by ion bombardment. In order to achieve high etch rates and smoothly etched surfaces a high plasma density, reasonably high bias and low pressure for strongly directional ion bombardment of the sample are required. For these reasons the etchings were done with a RIE100 (Oxford Instruments) ICP system. Optimal etching parameters were determined resulting to highly directional sidewalls and smooth surface of etched DLC structures. The same method was applied for the DLC:Ag nanocomposite containing 8 at.% of Ag. In this case during the $O_2$-RIE the DLC was effectively etched but the exposed Ag particles started acting as hard mask preventing the etching of the underlying DLC. In order to overcome this issue, combined $O_2$ and Ar plasma was used, resulting in a more efficient etching of the nanocomposite material. Although, the selectivity of the mask with respect to the composite material is slightly reduced it was still large enough to allow the patterning. Etch rates of both DLC and DLC:Ag against SiPol are presented in Fig. 3. Furthermore, these etch rates are compared to that of a commonly used resist such as PMMA to further validate them. After etching, the remaining residues of SiPol were removed in $CHF_3/CF_4$ plasma followed by rinsing with acetone and isopropanol. Care has to be taken that this descum using fluorinated plasma does not lead to the extensive formation of silver fluoride, which is likely to impair the plasmonic properties of the silver nanoparticles. This could e.g. be achieved by choosing an appropriate initial resist thickness by which the resist is almost completely removed by the combined $O_2$ and Ar plasma step. The prolonged exposure to solvents, e.g., acetone or PGMEA, can also provide sufficient removal of resist residues. Etching recipes for the different materials involved in processing are presented in Table 1. Finally, surface characterization was done using SEM (Zeiss Supra VP55).

**Table 1. Etching parameters.**

| Etch step | DLC | DLC:Ag | SiPol removal |
| --- | --- | --- | --- |
| Gas (sccm) | 10 $O_2$ | 10 $O_2$, 10 Ar | 10 $CHF_3$, 10 $CF_4$ |
| Power ICP (W) | 1000 | 1000 | 900 |
| Power bias (W) | 80 | 80 | 30 |
| Pressure (mTorr) | 11 | 11 | 11 |

2.5 Grating reflectance spectra modeling and optical characterization

Numerical simulations of the polarized white light reflection spectrum from the DLC grating were performed employing RCWA based GSolver software (Grating Solver Development Co.). In RCWA method the results are provided for one frequency only, therefore multiple simulations are necessary in order to cover the broadband spectrum. The profile relief and optical properties of the structure were chosen according to the SEM micrographs of the actual investigated samples (as seen in Fig. 1c). The grating was approximated by a finite number of slabs to the closest ideally rectangular profile. Additional information regarding numerical simulations can be found in [24].

Experimental optical measurements of GMR response in DLC linear gratings were carried out using custom optical setup consisting of a white light source (halogen lamp), collimating optics and a Glan–Taylor polarizing prism, thermostated fluidic cell with sensor chip, fiber optics and spectrometer (see Fig. 4). A fluidic cell together with the diffraction grating was mounted on a goniometric stage (resolution 1′). The fluidic cell is fabricated from a polytetrafluoroethylene (PTFE) based compound. On the front side of the cell, the chip with diffraction grating is fixed and on the back side a thermoelectric Peltier element. This thermoelectric element is water cooled and driven by TEC-1090 (Meerstetter Engineering GmbH) temperature controller allowing precise temperature setting and stability up to 0.01°C during the measurements. Temperature was monitored with PT100 temperature sensors. The cavity in between the thermoelectric element and the DLC sensor chip is filled with the liquid analyte through several inlets and outlets. As a proof of concept, different mass concentrations of isopropanol (IPA) solution in DI water (0-100%) were used as analyte. The refractive index of the different solutions (ranging from 1.330 for pure DI water to 1.375 for pure IPA) was determined using an Abbe refractometer (Carl Zeiss). Then the analyte was filled into the measurement cell with the DLC sensor chip to determine the optical response. All experiments were carried out at 23 °C temperature and 20% humidity. The measured spectra were normalized to the lamp spectrum taking into account different polarizations and integration times. The spectrum was collected at angle of incidence of 20º.

The polarized white light spectra reflected from grating was coupled to an optical fiber connected to a AvaSpec-2048 (Avantes) spectrometer using a quartz lens fixed on a platform that can be rotated independently around the same axis as the grating. The spectrometer collects the data with a spectral range between 360– 860 nm and a resolution of 1.2 nm. The whole setup is automated and controlled through custom LabView based software. The setup is described in more detail in [12,13].

## 3 Results and Discussion

### 3.1 Imprinted structures in Sipol resist

Due to poor adhesion between DLC and SiPol the conventional imprint routine when the resist completely fills stamp cavities was found to be unsuitable for densely packed patterns. Lack of adhesion between DLC and SiPol always resulted in complete or partial resist peel off from the substrate (silicon or DLC) during stamp separation. None of the usual measures, e.g. surface pre-treatment (plasma, baking), variation of imprint conditions or low demoulding temperature, provided significant positive effect. The combination of low-temperature and incomplete filling approach was found to be the best solution to cope with the issue and also greatly simplifies the process by enabling pattern transfer with the single etching step.

Imprinted structures into SiPol are presented in Fig. 5a-b. For the linear gratings the 70 nm high resist was equally distributed within the cavities and showed only low tendency to form capillary bridges (due to the high cavity height, the small lateral dimension and the low imprint temperature versus the $T_g$). Small rims at the sides of up ~20 nm are caused by wetting of sidewalls and do not interfere with a pronounced anisotropic etching (this even might reduce the rounding of the resist edge during etching). At the same time, almost zero residual layer thickness was achieved due to the high ability of the SiPol material to flow (Fig. 5a). Although, complete dewetting down to 0 nm seems possible, the actual remaining thickness is difficult to determine. However, even if a few nanometers remain, the following etching step (in $O_2$-plasma) will serve as an effective descum (window opening) step before the DLC is etched. In this case the remaining residual layer can be considered as effectively equal to zero. The same dewetting effect can also be seen in other type patterns, e.g. round holes of 300 nm in diameter ordered in hexagonal array (Fig. 5b). Small rims around the holes are due to wetting of sidewalls.

The maximum fill rate of stamp cavities for the linear grating patterns until delamination of SiPol from the underlying DLC substrate starts to occur was found to be at the aspect ratio of ~0.7. For not such dense patterns e.g., holes, where much larger area of resist is in contact with substrate than with stamp, this is less of an issue and the ratio could be higher.

### 3.2 Pure DLC etching with different patterns

The low-temperature, incomplete filling approach greatly simplifies the process by enabling the transfer of the masking layer pattern into DLC films with oxygen plasma. The etching of SiPol itself is a complex process which has two goals: the thinning down of the organic-inorganic polymer until windows are opened and the conversion of the remaining polymer into almost pure $SiO_2$ by removing the organic content. In the ideal case this results in a thin homogeneous layer of $SiO_2$ which would not be susceptible to heat in the extended oxygen process. Since we virtually achieved a zero residual layer patterning, a pure oxygen etching step was found sufficient to achieve the removal of any resist residues but this process would not convert the SiPol into $SiO_2$ completely, as for the standard SiPol removal process. Once the upper few nanometers of SiPol are converted, it would not further etch and develop into a thin skin on the still intact SiPol. The thermoplastic properties of this core would then be susceptible to any heat treatment, which in case of oxygen plasma was found enough that the resist lines would develop into thermally reflowed cylinder structures (see Fig. 1a). The reflow would be enhanced if undercuts develop which further pull the skin into a rounded shape (see Fig. 1b). This effect is greatly pronounced when SiPol is used in a bilayer system with purely organic material e.g., UL1. The advantage would be that we can create undercuts. However, this behavior is greatly diminished when SiPol is used directly on DLC. High etch rates prevents $SiO_2$ skin formation by a constant formation – erosion cycle during the etching and the mask thins down rather gradually.

Etching parameters for DLC were optimized to achieve highly directional transfer. From etching rates depicted in Fig 3. Selectivity between Sipol and DLC were determined to be at least 1:4.

The linear gratings and hexagonal structures (both holes and pillars) were successfully transferred into DLC. Fig. 6a-b shows 400 nm period 140 nm depth and 380 nm period 300 nm depth gratings etched into DLC. In Fig. 6c the 380 nm deep holes in DLC are not yet deep enough to etch through the entire, 450 thick film. At the same time, the SiPol is not yet completely removed (estimated 10-20 nm). It shows, in the area between the holes, a visible undulation, which is, as in the case of the linear gratings, due to the fact that the SiPol wets the stamp pillars. This is due to capillary effects (electro-hydrodynamic effect) and therefore shows some shallow voids and lateral displacement of material. The choice of process parameters thickness contrast shows that even in this special case no capillary bridges could form and that the surface undulations presenting the onset of capillary action were far from causing problems in etching.

The pillars in the DLC shown in Fig. 6d are completely etched, i.e. the 400 nm thick film is etched down to the silicon substrate and the SiPol removed by using an additional $CHF_3/CF_4$ step. In this case silicon substrate was also etched by fluorine species proving the use of DLC as an etching mask with a high selectivity towards silicon and silicone oxide.

3.3 Patterning of DLC:Ag composite films

The SiPol mask was also used to etch DLC films doped with silver (DLC:Ag). However, in this case etching with pure oxygen plasma wasn't sufficient, due to presence of silver nanoparticles scattered all over the surface (and in the bulk) with each of one acting as individual etching mask. In this case the addition of Argon in the plasma was beneficial resulting in more effective removal of silver.

Argon also contributed to the faster deterioration of SiPol resulting to lover selectivity of 1:2.5 towards DLC:Ag. The residues of SiPol on the surface were removed as in the pure DLC case. This process was employed to structure DLC:Ag films with 400 nm period and approximately 150 nm depth (the precise depth is hard to determine due to roughness of the etched profile) grating patterns (see Fig.7a) leaving a high concentration of silver nanoparticles on the surface (see Fig.7b). The proposed process allowed obtaining DLC gratings and simultaneously exposing the Ag nanoparticles to the environment.

3.4 Optical characterization of grating structures

Using the fabrication method described above, 380-400 nm period grating type structures were fabricated in DLC and their optical response was characterized. As can be seen in reflectance spectra (Fig. 8) there is a clear resonant peak at 669 nm which red shifts towards longer wavelengths as the IPA concentration in analyte increases as so refractive index. The simulated response (Fig. 8a) is in good agreement with the experimental one (Fig. 8b). Small differences, such as the broadening of the peaks, arise due to limitations of measurement system and from small imperfection in the fabricated structures, despite the high quality of the fabricated gratings.

A total number of hundred reflectance spectra measurements for each concentration were carried out and the automatically defined average value of the peak position was chosen for refractometry calibration plot. In (Fig. 9) the peak shift is depicted as a function of refractive index that was obtained from the Abbe refractometer, which follows a linear behavior as expected. From this graph, a sensor sensitivity of 319 nm/RIU was determined. This result represent an improvement close to 450% when compared with previously reported sensors DLC based on similar DLC films doped with silicon oxide and patterned using interference lithography [12, 24]. Such improvements are mostly due to the precise geometry control by the fabrication method and the defined optical properties of the material used. Furthermore, such sensitivity is much higher than reported recently for conventional grating sensors [26-28] and compares to that what can be achieved theoretically in high complexity resonators [29]. The detection limit of this sensor, calculated according methodology presented in [30], was determined to be $7.5 \times 10^{-5}$ RIU. However, this is limited mostly by the spectrometer resolution and not by the developed arrays. Broad resonant peak linewidth would also impose limitations on sensor performance when attempting to monitor very small changes close to detection limit. However, the high sensitivity of DLC based sensors enables utilization in areas where extreme sensitivity isn't necessary but rather cost and robustness are the key factors providing high accuracy with less sophisticated peripherals, i.e. spectrometers, etc. The use of T-NIL combined with the SiPol resist has allowed to straightforwardly pattern complex materials that usually will require challenging and costly processes. Enabling the manufacturing of novel DLC based systems with enhanced capabilities in a cost-efficient and reproducible mass-manufacturing method.

## 4  Conclusions

We demonstrated the successful use of the new thermoplastic resist SiPol as a hard mask to pattern DLC films by thermal NIL process. SiPol can serve as efficient hard mask for etching DLC coatings, ensuring 1:4 selectivity for non-doped and 1:2.5 for silver doped DLC coatings. Adhesion issues can be addressed by using low temperature imprint in combination with incomplete filling of stamp cavities. The low temperature imprint approach, combined with incomplete filling of stamp cavities did not only allow a virtually zero residual layer imprint but also facilitated the original SiPol process involving an organic transfer layer by removing intermediate etchings resulting into minimum number of processing steps. DLC coatings and to some extent DLC coatings doped with silver can be effectively etched with oxygen plasma (ICP) with high precision and vertical sidewalls at least up to 500 nm depth. This enables the homogeneous transfer of relatively large area patterns of $20 \times 20$ mm$^2$ into DLC. The current applications have moderate aspect ratio and resolutions down to several hundred nanometers, which does not seem to be the final limit. High quality gratings fabricated by this method have been characterized as DLC leaky-waveguides-based sensors. The optical response shows a high agreement with simulations of ideally rectangular gratings, resulting in a measured high sensitivity (319 nm/RIU) to refractive index changes in the analyte. This represents an improvement close to 450 % when compared to previous works. These results confirm the high quality of the obtained structures, validating the use thermal NIL for patterning of DLC based devices with a high degree of reproducibility.

**Acknowledgments:**

This work was realized within the project and funded by the Scientific Exchange Programme SCIEX-NMS (Project n° 13.193) between Switzerland and the New Member States of the European Union. Additional partial financial support came from the Swiss National Science Foundation (SNF) Ambizione project (n° PZ00P2_142511) granted to V.J.C. and the project NFFA (n° 654360) under the H2020 European research and innovation programme. K. Vogelsang, A. Weber, A. Vasiliauskas, and V. Kopustinskas are acknowledged to for their technical assistance.

**References**

[1] J. Taniguchi, Y. Tokano, I. Miyamoto, M. Komuro, M. Komuro and H. Hiroshima, Diamond nanoimprint technology, Nanotechnology 13, (2002), 592–596.

[2] B. Nöhammer, J. Hoszowska, A. K. Freund, C. David, Diamond planar refractive lenses for third- and fourth-generation X-ray sources, J. Synchrotron Radiat. **10**(2) (2003) 168–171.

[3] P. Forsberg, M. Karlsson, High aspect ratio optical gratings in diamond, Diam. Relat. Mater. 34 (2013) 19–24.

[4] J. Robertson, Diamond-like amorphous carbon, Mater. Sci. Eng. R Reports. 37 (2002) 129–281.

[5] A. Grill, Diamond-like carbon: state of the art, Diam. Relat. Mater. 8 (1999) 428–434.

[6] H. J. Lee, B.S. Kwon, Y. R. Park et al., Inductively coupled plasma etching of chemical-vapor-deposited amorphous carbon in $N_2/H_2/Ar$ chemistries, Journal of the Korean Physical Society **56** (2010) 1441-1445.

[7] J.K. Kim, S. Il Cho, N.G. Kim, M.S. Jhon, K.S. Min, C.K. Kim, et al., Study on the etching characteristics of amorphous carbon layer in oxygen plasma with carbonyl sulfide, J. Vac. Sci. Technol. A Vacuum, Surfaces, Film. 31 (2013) 021301.

[8] K.A. Pears, A new etching chemistry for carbon hard mask structures, Microelectron. Eng. 77 (2005) 255–262.

[9] K.A. Pears, J. Stolze, Carbon etching with a high density plasma etcher, Microelectron. Eng. 81 (2005) 7–14.

[10]   M. Kakuchi, M. Hikita, T. Tamamura, Amorphous carbon films as resist masks with high reactive ion etching resistance for nanometer lithography, Appl. Phys. Lett. 48 (1986) 835-837.

[11]   S. Ramachandran, L. Tao, T.H. Lee, S. Sant, L.J. Overzet, M.J. Goeckner, et al., Deposition and patterning of diamondlike carbon as antiwear nanoimprint templates, J. Vac. Sci. Technol. B Microelectron. Nanom. Struct. 24 (2006) 2993–2997.

[12]   T. Tamulevičius, R. Šeperys, M. Andrulevičius, V. Kopustinskas, Š. Meškinis, S. Tamulevičius, et al., Application of holographic sub-wavelength diffraction gratings for monitoring of kinetics of bioprocesses, Appl. Surf. Sci. 258 (2012) 9292–9296.

[13]   T. Juknius, T. Tamulevičius, I. Gražulevičiūtė, I. Klimienė, A.P. Matusevičius, S. Tamulevičius, In-situ measurements of bacteria resistance to antimicrobial agents employing leaky mode sub-wavelength diffraction grating, Sensors Actuators B Chem. 204 (2014) 799–806

[14]   J.K. Luo, Y.Q. Fu, J. a. Williams, S.M. Spearing, W.I. Milne, Diamond and diamond-like carbon MEMS, J. Micromechanics Microengineering. 17 (2007) 147–163.

[15]   D. Threm, Y. Nazirizadeh, and M. Gerken, Photonic crystal biosensors towards on-chip integration, J Biophoton. 5(8-9) (2012) 601–616.

[16]   J.M. Jaramillo, R.D. Mansano, L.S. Zambom, M. Massi, H.S. Maciel, Wet etching of hydrogenated amorphous carbon films, Diam. Relat. Mater. 10 (2001) 976–979.

[17]   Y. Sasaki, A. Takeda, K. Ii, S. Ohshio, H. Akasaka, M. Nakano, et al., Evaluation of etching on amorphous carbon films in nitric acid, Diam. Relat. Mater. 24 (2012) 104–106.

[18]   T. Tamulevičius, A. Tamulevičienė, D. Virganavičius et al., Nucl. Instrum. Meth. B 341 (2014) 1-6.

[19]   H. Schift, Nanoimprint Lithography: An old story in modern times? A review, J. Vac. Sci. Technol. B 26(2) (2008) 458-480.

[20]   M. Messerschmidt, A. Schleunitz, C. Spreu, T. Werner, M. Vogler, F. Reuther, et al., Thermal nanoimprint resist for the fabrication of high-aspect-ratio patterns, Microelectron. Eng. 98 (2012) 107–111.

[21] A. Tamuleviciene, S. Meskinis, V. Kopustinskas, S. Tamulevicius, Multilayer amorphous hydrogenated carbon (a-C:H) and SiO(x) doped a-C:H films for optical applications, Thin Solid Films. 519 (2011) 4004–4007.

[22] Š. Meškinis, A. Čiegis, A. Vasiliauskas, A. Tamulevičienė, K. Šlapikas, R. Juškėnas, et al., Plasmonic properties of silver nanoparticles embedded in diamond like carbon films: Influence of structure and composition, Appl. Surf. Sci. 317 (2014) 1041–1046.

[23] S.Y. Chou, P.R. Krauss, P.J. Renstrom, Imprint of sub-25 nm vias and trenches in polymers, Appl. Phys. Lett. 67, 3114–3116 (1995)

[24] T. Tamulevicius, I. Gražuleviute, D. Urbonas, M. Gabalis, R. Petruškevicius, S. Tamulevicius, Numerical and experimental analysis of optical response of sub-wavelength period structure in carbonaceous film for refractive index sensing, Opt. Express. 22 (2014) 27462–27475.

[25] B.S. Kwon, J.S. Kim, N.-E. Lee, J.W. Shon, Ultrahigh selective etching of $SiO_2$ using an amorphous carbon mask in dual-frequency capacitively coupled $C_4F_8$/$CH_2F_2$/$O_2$/Ar plasmas, J. Electrochem. Soc. 157 (2010) D135.

[26] A. Szeghalmi, E.B. Kley, M. Knez, Theoretical and experimental analysis of the sensitivity of guided mode resonance sensors, J. Phys. Chem. C. 114 (2010) 21150–21157.

[27] G. Hermannsson, K.T. Sørensen, C. Vannahme, C.L.C. Smith, J.J. Klein, M. Russew, et al., All-polymer photonic crystal slab sensor, Opt. Express. 23 (2015) 16529–16539.

[28] Y.-F. Ku, H.-Y. Li, W.-H. Hsieh, L.-K. Chau, G.-E. Chang, Enhanced sensitivity in injection-molded guided-mode-resonance sensors via low-index cavity layers, Opt. Express. 23 (2015) 14850–14859.

[29] S. Abbas, J. Moghaddas, M. Shahabadi, S. Member, M. Mohammad-Taheri, Surface sensitivity using coupled cross-stacked gratings, IEEE Sensors Journal, 14 (2014) 1216–1222.

[30] I.M. White, X. Fan, On the performance quantification of resonant refractive index sensors, Opt. Express.16 (2008) 1020–1028.

**Figures and Tables**

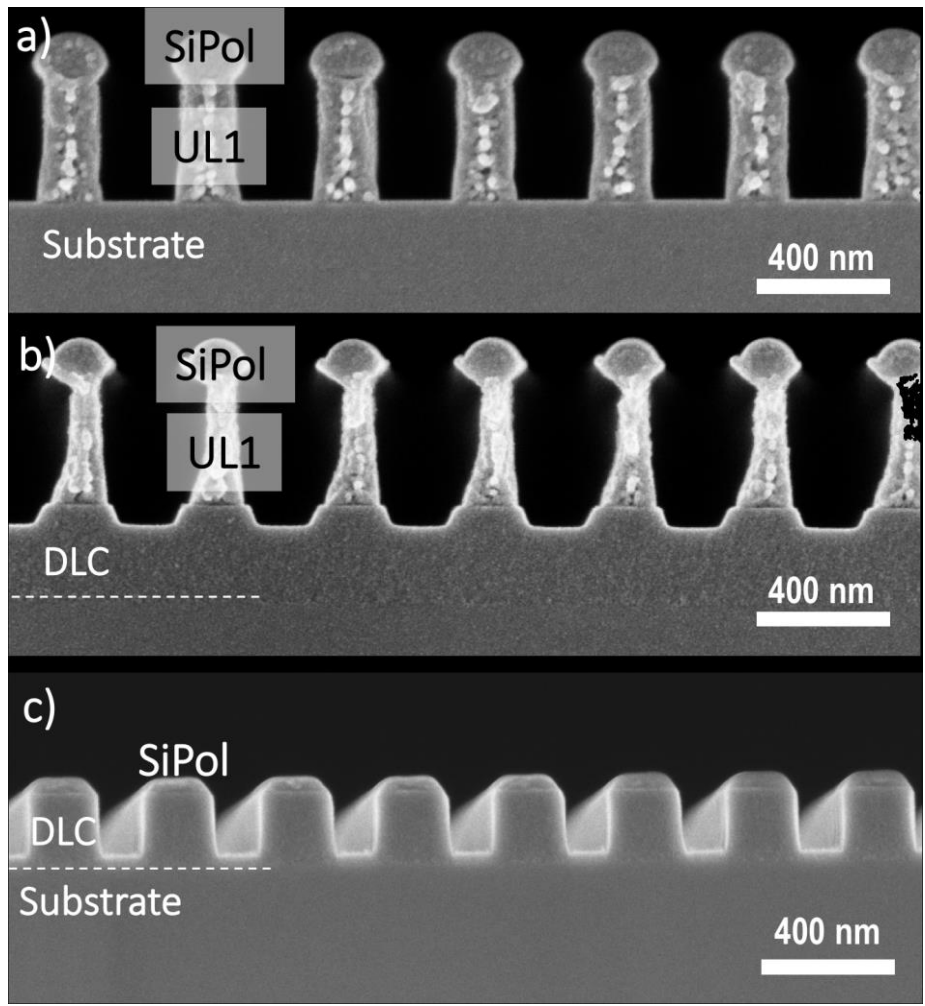

Figure 1: Cross-sections of etched grating structures A) Structures etched in bi-layer SiPol-UL1 resist mask B) Bilayer SiPol-UL1 resist mask used to etch into DLC showing high under etching of UL1 layer. C) DLC etched using only SiPol resist mask

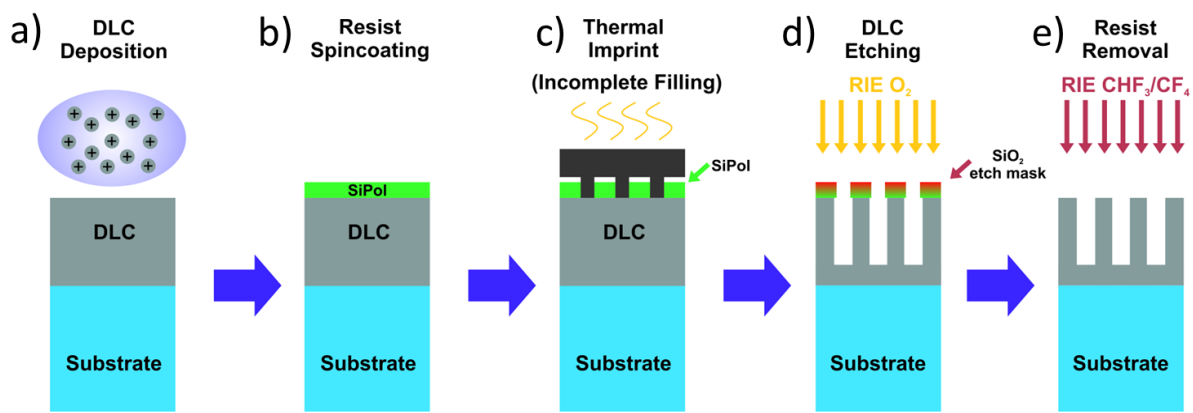

Figure 2: Schematic depiction of processing steps of DLC patterning using SiPol resist as hard mask. a) Plasma enhanced chemical vapor deposition process was used to grow thin DLC films. b) Thin layer of silicon containing resist was spin coated directly on DLC. c) Nanoimprint lithography with incomplete stamp filling was used to replicate reverse patterns of the silicon stamp into resist layer. d) Reactive ion etching by oxygen plasma was employed to transfer patterns into DLC. e) Reactive ion etching by fluorine plasma was employed to remove remaining resist layer of DLC surface.

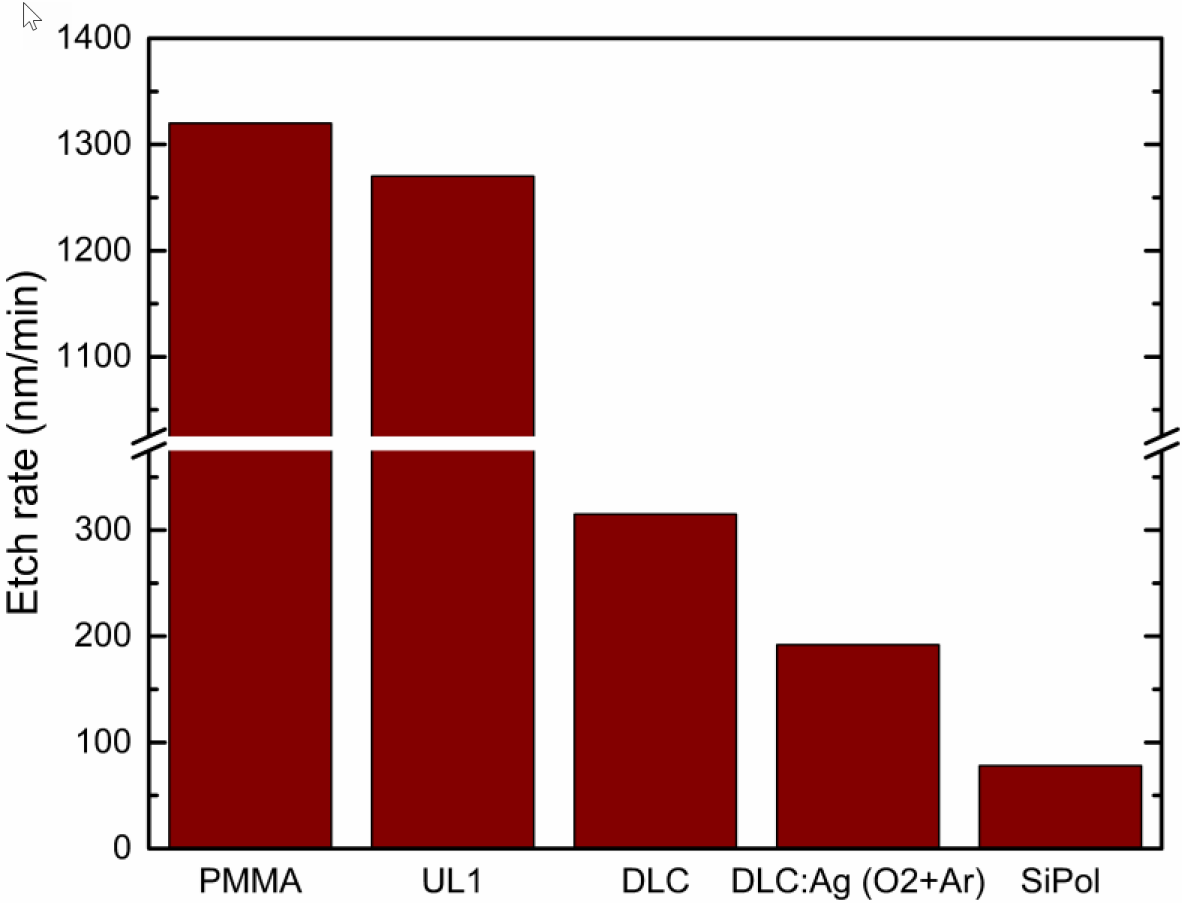

Figure 3: Etching rates in the oxygen plasma at the optimized parameters for DLC etching.

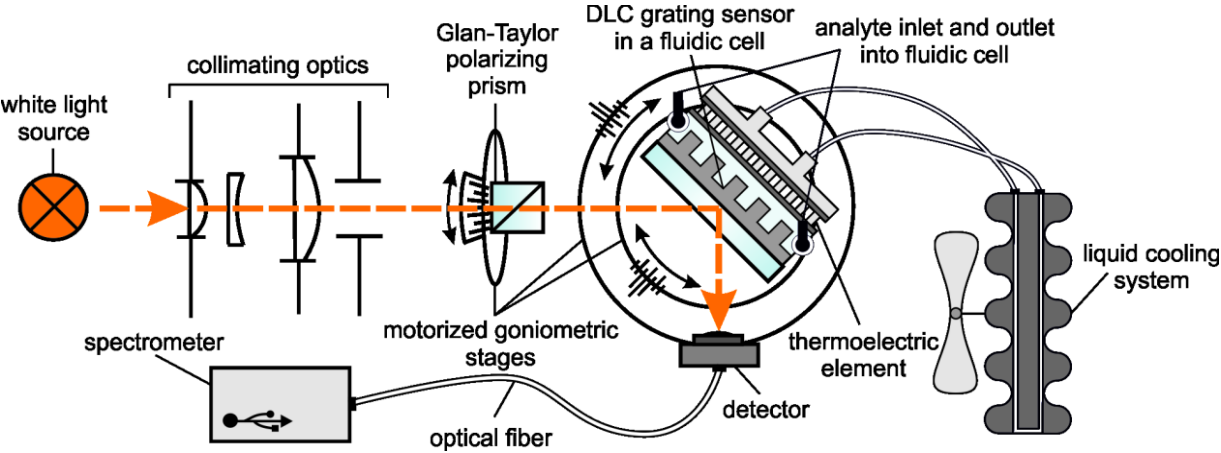

Figure 4: Schematics of optical setup hosting the DLC grating on a transparent substrate in a fluidic cell.

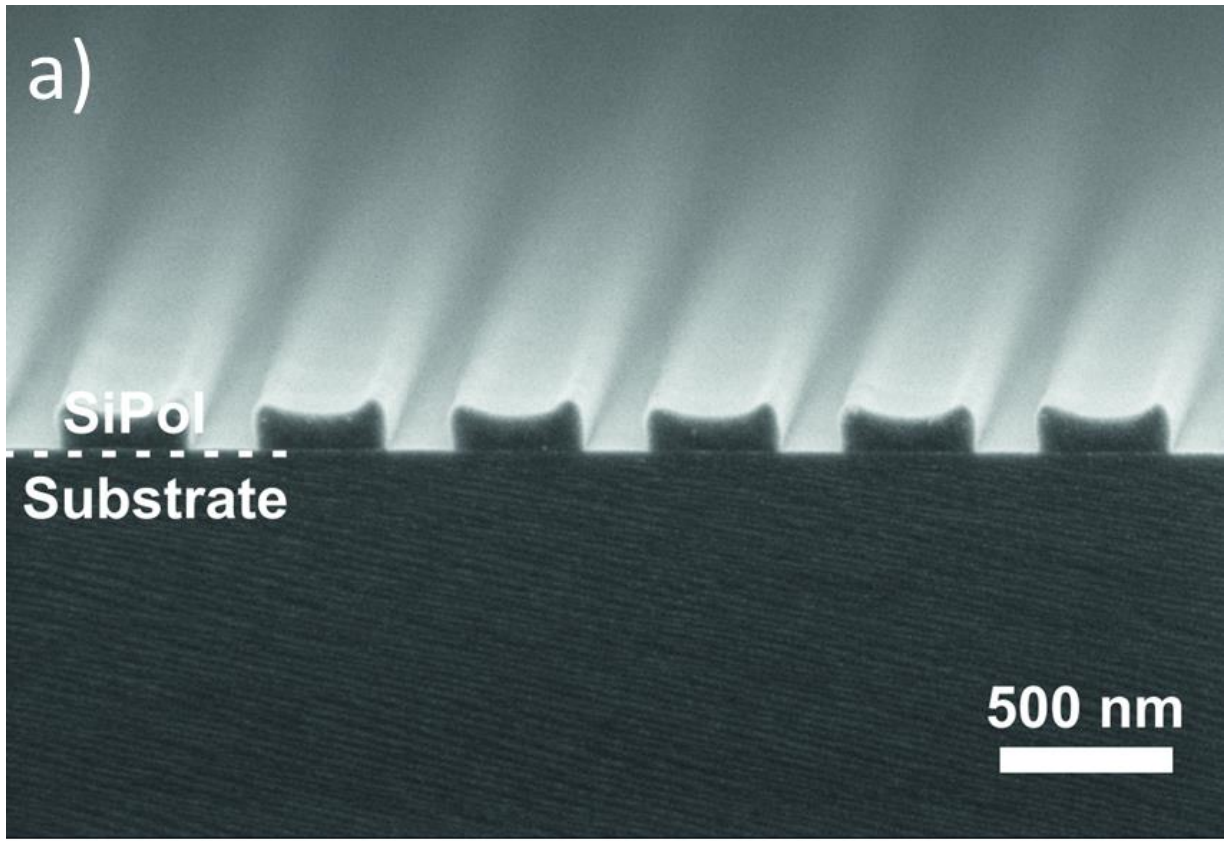
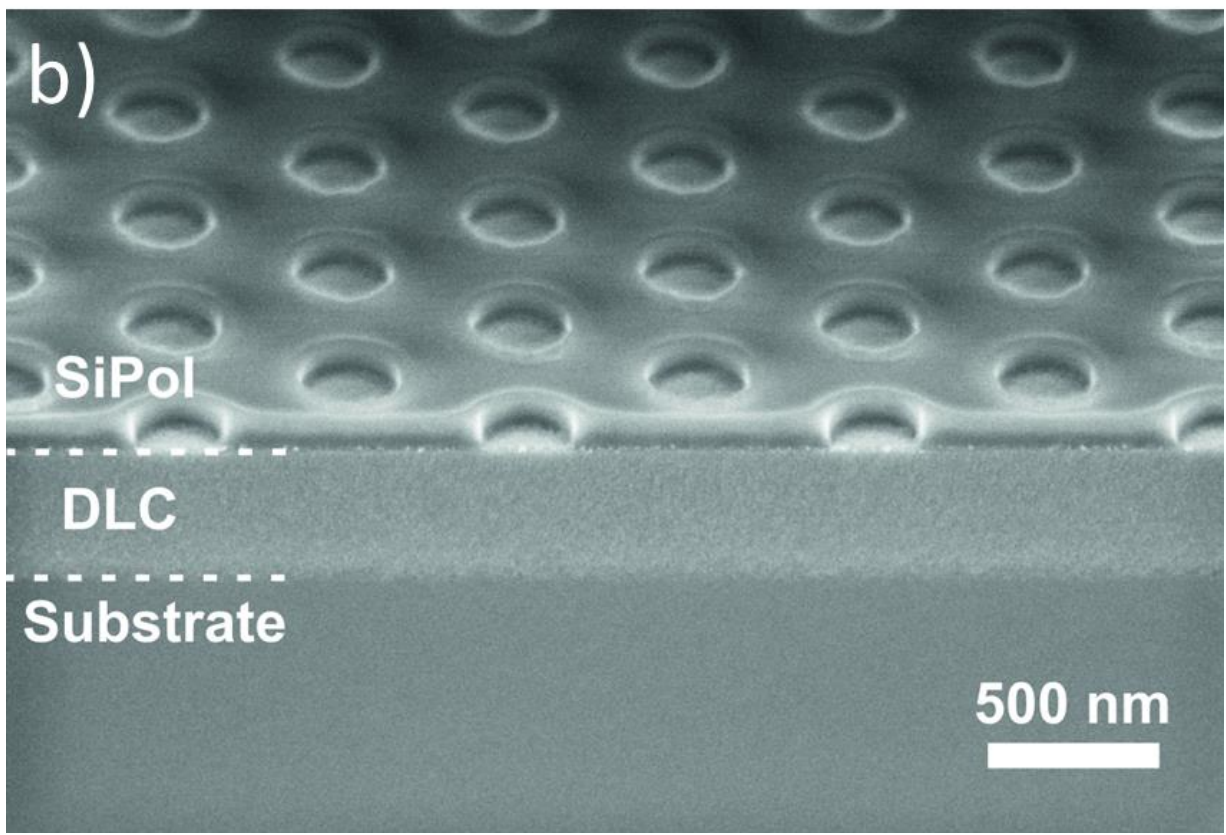

Figure 5: Patterning of SiPol with incomplete filing strategy: a) imprinted 580 nm period grating b) imprinted 300nm diameter holes ordered in hexagonal array.

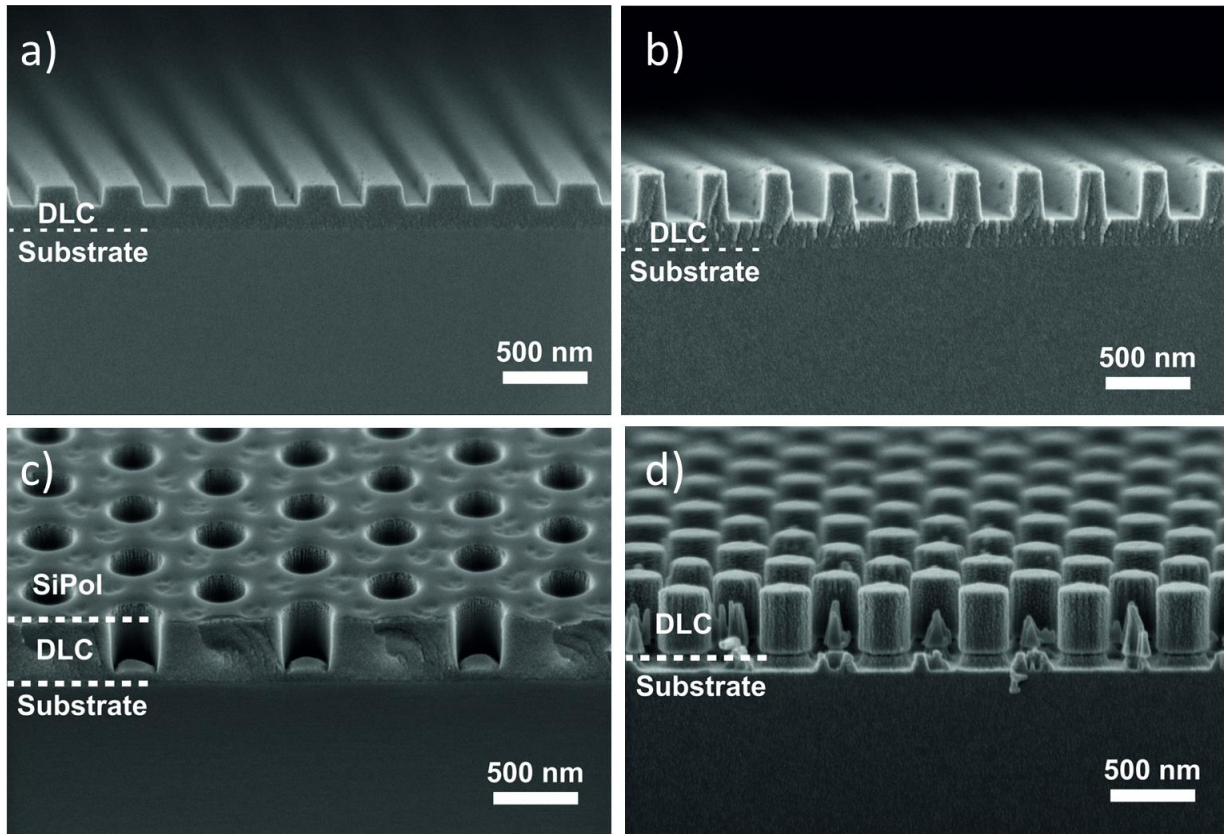

Figure 6: Patterning DLC: a) 400 nm period 140 nm depth grating etched into DLC b) 380 nm period 300 nm depth grating etched into DLC c) hexagonal hole pattern (300 nm diameter, 380 nm depth) etched into DLC, d) hexagonal pillars pattern (300 nm diameter, 400 nm depth) etched into DLC and substrate.

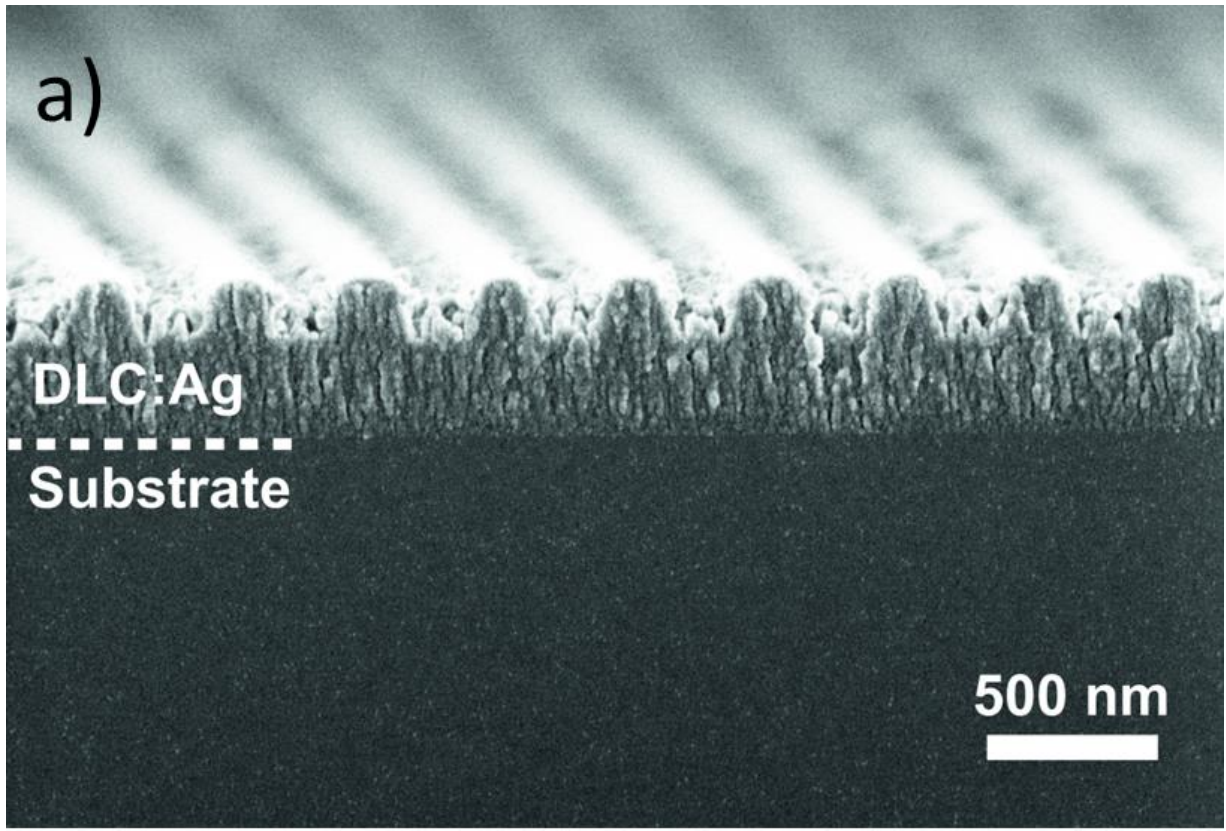
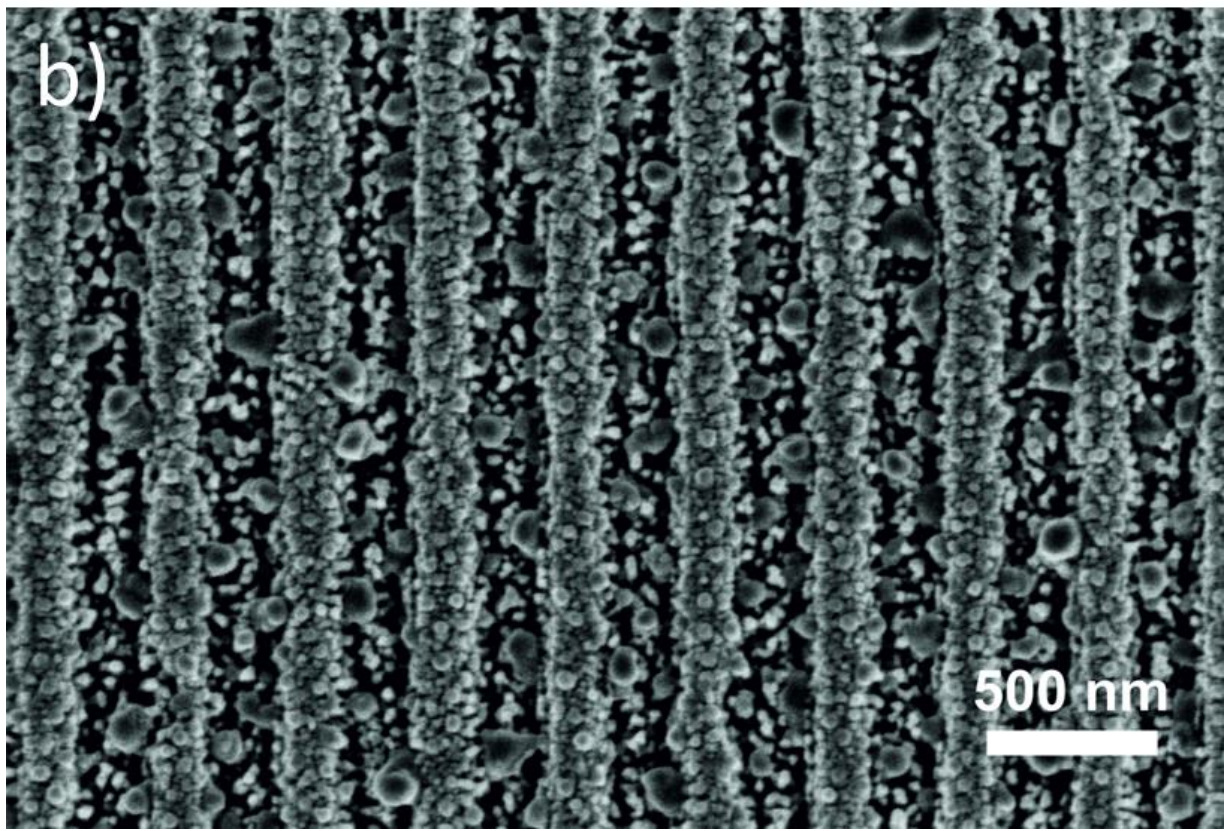

Figure 7: 400nm period and 150nm depth gratings in DLC:Ag nanocomposite a) from side, b) from top

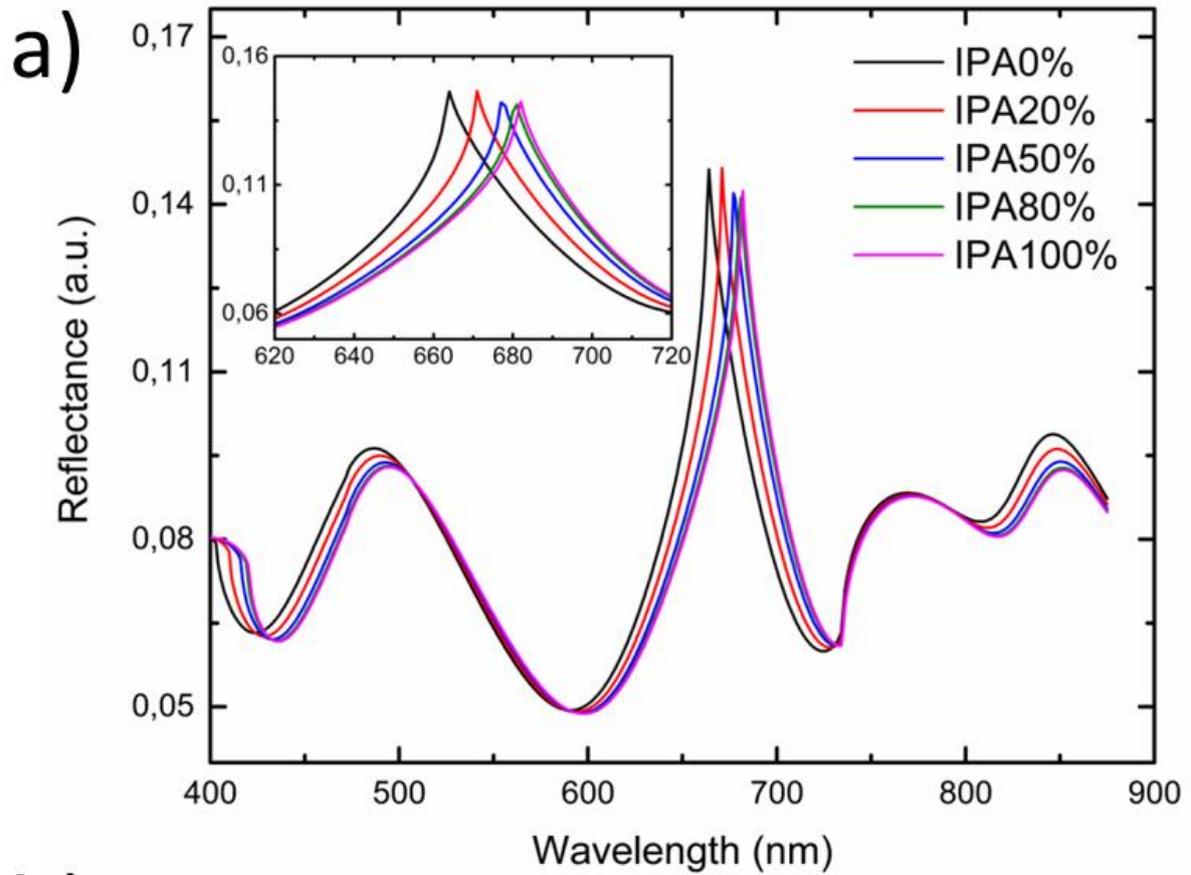

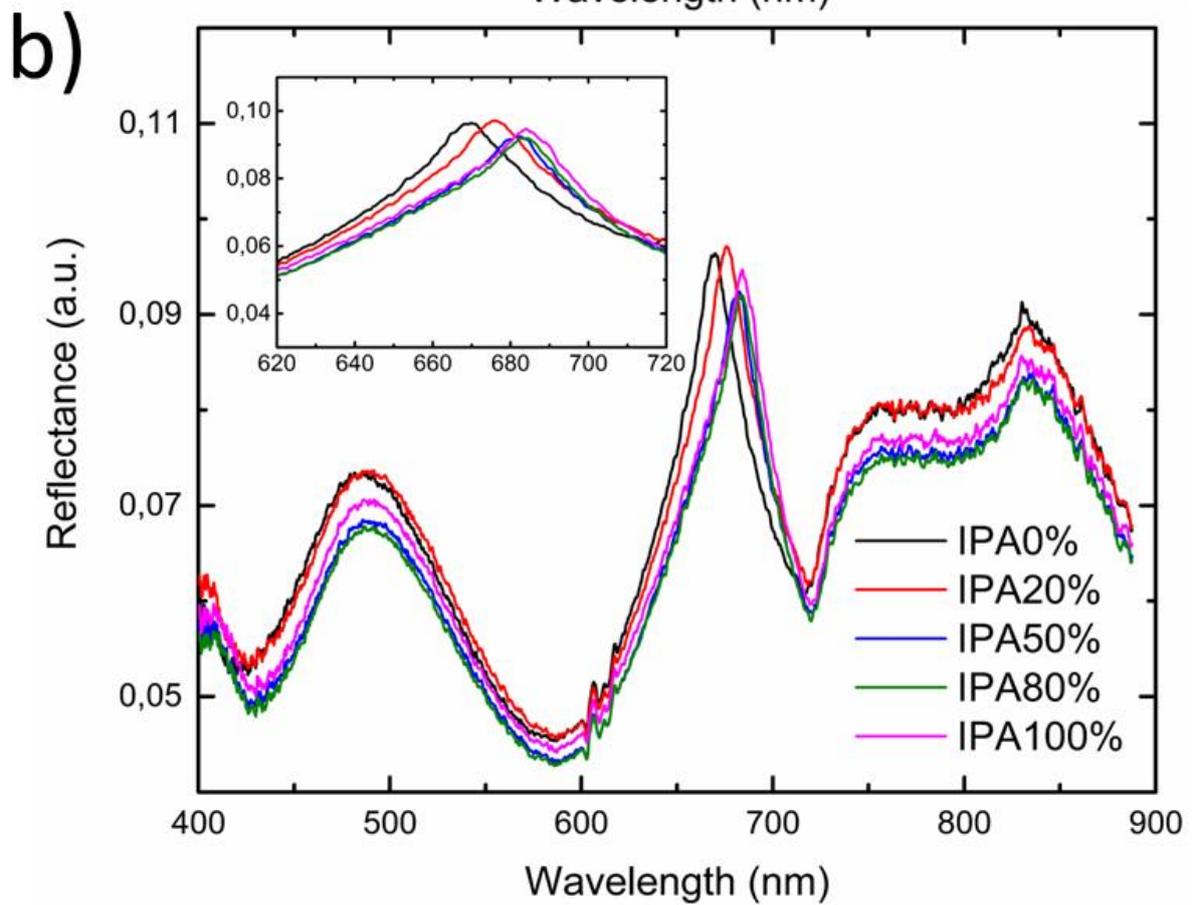

Figure 8: Simulated (a) and measured (b) reflectance spectra of different concentration of isopropanol

solution in DI water showing GMR peak shift due to changing refractive index. Measured and simulated with TM polarization at 20º angle of incidence of 400 nm period and 220 nm depth with 40 nm waveguide layer grating fabricated in DLC. Insets show zoomed graphs at the resonance peaks.

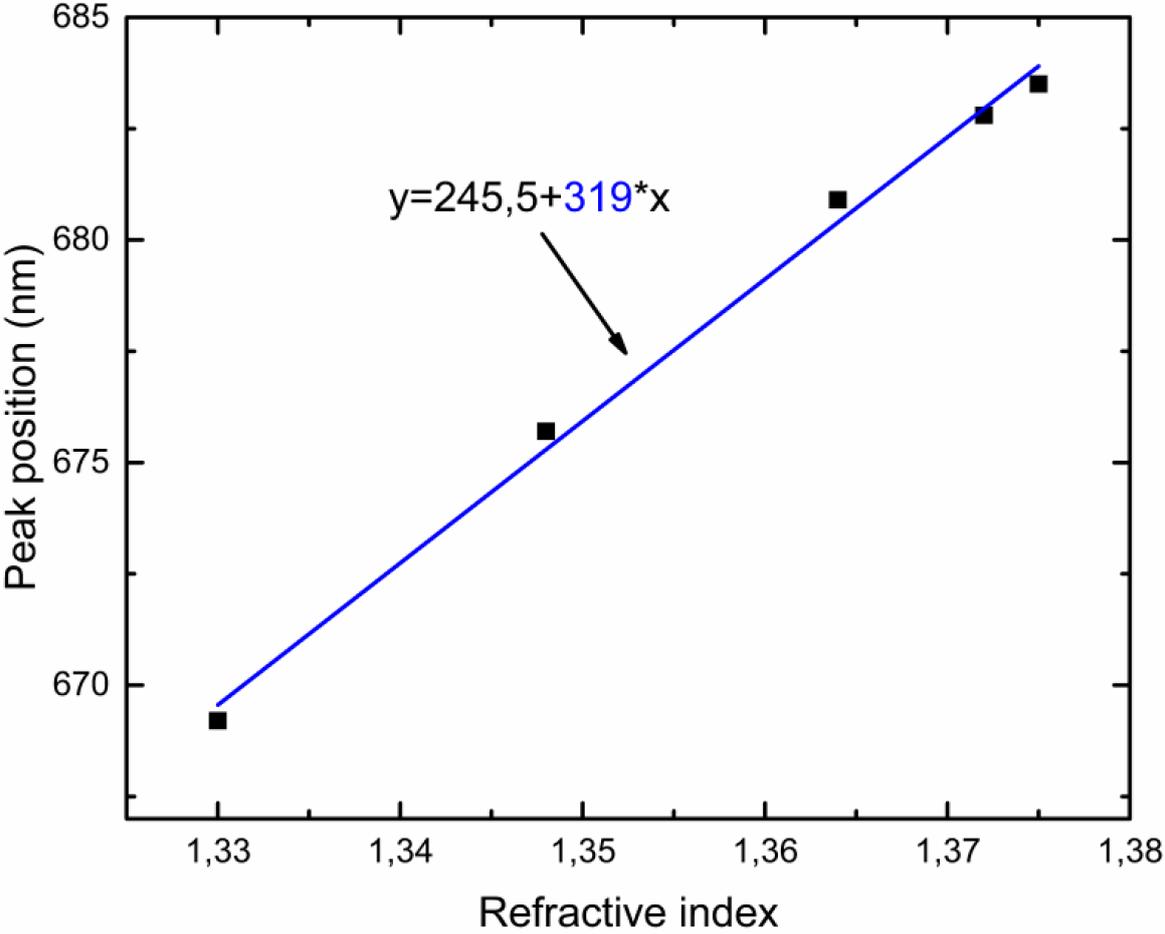

Figure 9: Experimentally obtained resonance peak position shift as a function of the analyte refractive index.